# Remark on Repo and Options

Andrei A. Kapaev[1]

[1]N.N.Bogolubov Lab of Geometric Methods in Mathematical Physics, Mechanic and Mathematics Department, Moscow State University, Leninskie gory, 1, Moscow, 119991, Russian Federation
Corresponding Author

**Abstract** The general and special repo rates are related with the prices of the European call- and American put-options. The evaluation takes into account specific business models of the parties in the repo agreement and the law restrictions. Using the repo-option relation, an alternative to the Black-Scholes method of option pricing is presented. The empirical data on the general and special repo rates are explained.

## 1. Introduction

Here, we are interesting in a class of derivatives with an enormous market volume known as the repurchase agreement, or repo [7]. The repo market is one of the important suppliers of the liquidity for bank institutions, finance corporations etc. while the central bank repo rate is one of the most significant finance indicators.

In principal, the repurchase agreement is a collateral loan, and virtually any portfolio of securities can be used as the collateral. The available data on the repo rates usually concern the Treasuries repo, i.e. the repurchase agreements with a portfolio of Treasuries accepted as the collateral. Usually, under the repo agreement, borrowers find a short term supply of the least expensive liquidity, while lenders find the least risky investments of their funds. The described business model corresponds to the so-called *general* repurchase agreement, and the corresponding interest rate is called the *general* repo rate.

The scheme of the repurchase agreement supports also other business models where the lender makes profit via short-term investments of the borrowed securities. Namely, if an investor expects a drop in the price of the securities, he might borrow the securities using the repo agreement and maintain a short position in the market. This kind of repo is called *special*, and the respective repo interest rate is called the *special* repo rate.

The differences in the general and special repo rates are studied e.g. in [8]. A connection of the special repo and options was observed in [6] where the authors discuss strategic "fails" and explain differences between the rates in some particular forms of the special repo. In the empirical study of the term structure of the repo rate on the three separate repo markets based on the Treasuries, the agencies securities and mortgage-backed securities as collateral [12], it was shown that the measured variation in the repo rate comes from the collateral risk and is irrelevant to the counterparty credit risk. This is an important result for our model explained below.

We also mention theoretical papers on various aspects of the repurchase agreement theory like the coexistence of the repo and direct sales [10], search price model [14] and the special repo rates [2].

The classical option pricing model by Black and Scholes [1], thanks to its relative simplicity and flexibility, is adapted to any kind of options and other derivatives and became a practical tool for many traders, hedgers and investors. It is widely believed that this model helped the option market to reach an extraordinary volume even though the Black-Scholes model predictions do not match the empirical market data [1], not saying it uses several non-realistic assumptions. Probably, this influence reflects mainly the people's need to have a simple and flexible tool to navigate in the uncertain market environment rather than the actual efficiency of the model in decision making [15].

Below, we connect the problem of the repo rate evaluation with the problem of the option pricing. We present an alternative to the Black-Scholes method of evaluation of the call-option price implied by the general repo agreement. We observe that our approach produces numeric results remarkably close to the figures computed using the Black-Scholes formula. We also relate the problem of the American put-option pricing with the evaluation of the special repo rate.

## 2. Methods

By definition, the interest rate is the rate at which a borrower pays a fee for the use of money owned by a lender. The variety of existing interest or discount rates is rather large and includes those on retail and corporate loans, on bonds and notes, on municipal and government securities, on reserves and excess reserves etc. The values of the interest and discount rates depend on the form of the borrowed asset, credit score of the borrower, loan duration and other aspects including overall economic conditions.

Among the most important interest rates, we mention the discount rate of the Federal Reserve to depository institutions on overnight primary credits (recently, 0.75%), seasonal credits (0.15%) and on Federal Funds (0-0.25%). All these loans are fully collateralized.

Another class of important interest rates is the yields of Treasuries. This year, the latter vary from 0.01% for 1 month bills up to 3.6% for 30-year bonds.

Finally we mention that this June the overnight Treasury general repo rate dropped to 0.12% that is lower than the Fed interest rate on seasonal loans for small and medium banks.

Looking for simplicity (cf. [1]), in our subsequent analysis, we assume the perfect liquidity of the market and the absence of any transaction cost. We also assume that the parties of the transactions interpret the conditions of the agreement according to the prescribed business models and the possibilities provided by the market conditions. Formulating the business model and behavior of the participants, we mainly follow the study of the repo market by M. J.Fleming and K. D. Garbade [6].

Namely, the general repo agreements are interpreted as the collateral loans, and the decisions of the general repo parties closely follow the conventions of such kind of loans.

Following [6], we distinguish the special repurchase agreement under the Federal Reserve supervision and the special repo between private corporations. In the first case, the specific Treasures demanded by the market are supplied by the Fed for a *fee* in the form of "bonds-versus-bonds" loan. Therefore the relevant repo rate is negative. In the second case, the private corporations delivering the specific securities, in contrast to the Fed, might fail not delivering the securities one or more days, and the corresponding special repo rate can approach the zero value but never becomes negative [6].

Theoretically, there is a possibility that the borrower of securities can fail not returning the borrowed specific securities at the agreed day. However, we were unable to find the empirical market data which could agree with the predictions of the corresponding model.

The general idea underlying our analysis is the fact that any obligation assumed by an agreement but not supported by an appropriate contractual commitment can be avoided by either party. From the finance point of view, this fact manifests itself as if the party that interprets the *obligation* as the *right* has purchased an option written by the other party. Introduction of the options in the analysis of the typical business interactions allows us to describe their finance side in clear terms, e.g. to connect the repo rates with the option prices, avoiding a difficult study of random and chaotic processes, see e.g. [3]. Comparing the results achieved in this way with the empirical market data, it is possible to rule out unrealistic models and confirm other ones.

## 3. Repurchase Agreement

A financial arrangement is called repo (repurchase agreement) if a holder of securities sells them to a lender and agrees to repurchase them at an agreed future moment at an agreed price.

Signing the repo agreement, the lender accepts a collateral risk while the borrower accepts the risk of not receiving its securities back at the closing leg since both parties retain the opportunity to "fail". The lender even accepts the risk of not receiving the agreed collateral [6]. It is important that the default of either party of the repo agreement does not imply an appeal to a bankruptcy court. We also assume that the "fails" do not incur ancillary costs like capital charges [6].

In spite of the risk of default, the use of repo is massive [7] because it serves as a primary source of the least expensive liquidity in the financial markets.

Sometimes, the demand on specific securities drives the repo rates to negative levels, see e.g. [6, 7]. The latter rates associated with the borrowing of securities instead of money are called special to distinguish them from the former general repo rates.

To avoid all the unnecessary complications, we restrict our considerations to the one-period model. At the starting leg, all prices are known constants, while the forward prices and returns at the closing leg are understood as the random variables with the particular expectations and deviations.

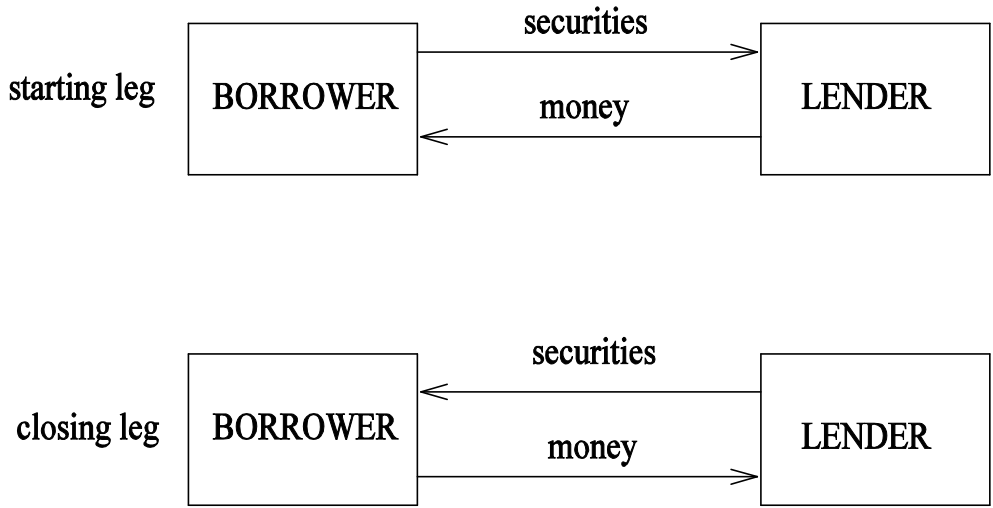

Figure 1. Scheme of the repurchase agreement.

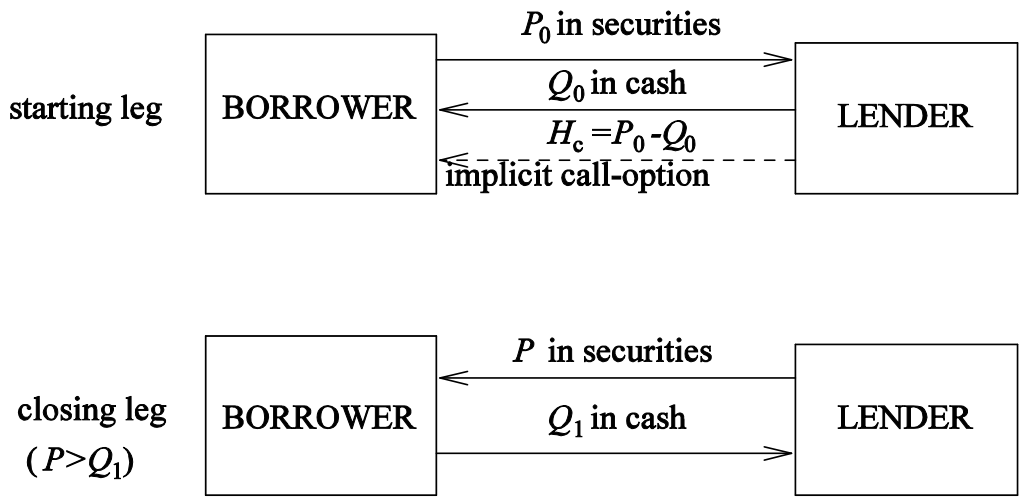

implicit call-option is executed

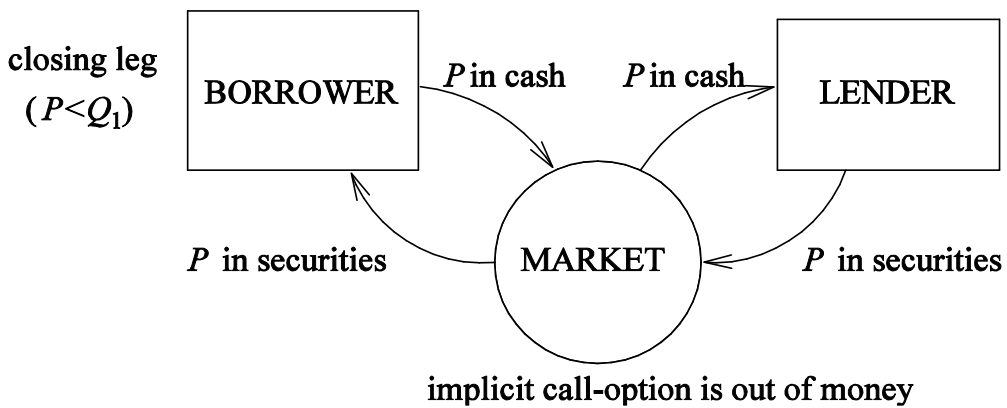

implicit call-option is out of money

Figure 2. The transaction scheme of the general repo agreement.

### 3.1. General Repo

We are going to re-interpret the economic processes underlying the general repo agreement using the assumption that the behavioral patterns of both principal participants of the agreement follow particular models.

The general repurchase agreement is initiated by the borrower of money who owns some securities in possession and looks for a short-term liquidity. The primary lender's business is crediting, and the securities provided by the borrower serve as the collateral for the loan.

The model of the collateral loan implies that the lender has no motivation to break the agreement while the borrower *might claim a default* under particular conditions. Such a situation admits an interpretation as if the borrower buys a European call-option for the securities provided as collateral.

In formal terms, our convention on the general repurchase agreement is the following.

The borrower of funds:

1). at the starting leg, delivers the securities worth $P_0$ to the lender and receives $Q_0 = P_0(1 - h_c)$ in cash, where the discount $H_c = P_0 h_c$ is called *haircut* (we call $h_c$ the haircut rate);

2). simultaneously, the borrower receives from the lender the *implicit* call-option at the present price $H_c$ and the strike price $Q_1$;

3). at the closing leg, if the market price of the securities becomes higher than the call-option strike price, i.e. $P > Q_1$, then the borrower pays $Q_1 > Q_0$ to the lender for securities worth $P$ (the implied call-option is exercised);

4). at the same closing leg, if the market price of the securities is lower than the call-option strike price, i.e. $P < Q_1$, then the borrower buys the securities at $P$ in the market (the implicit call-option is out of money).

The lender of funds:

1). at the starting leg, lends $Q_0$ in cash receiving the securities of value $P_0 > Q_0$;

2). simultaneously, sells at $H_c$ the implicit call-option for the same securities with the strike price $Q_1$;

3). at the closing leg, if the market price of the securities is higher than the strike price of the call-option, $P > Q_1$, the lender receives $Q_1$ in cash and delivers the securities to the borrower of funds (the implied call-option is exercised);

4). if the market price of the securities is less than the strike price of the call-option, $P < Q_1$, the lender sells the securities at the spot price $P$ (the implicit call-option is out of money).

Because the final result of all the transactions above depends on the forward price of the securities at the closing day and therefore contains an uncertainty, the conventional definition of the repo rate $r_R$,

$$r_R = \frac{Q_1}{Q_0} - 1, \quad (1)$$

does not reflect particularly important features of the agreement. Besides the general repo rate $r_R$, we introduce below the intrinsic yield $r_S$ of the collateral, the lender's interest rate $r_L$ and the implicit call-option yield $r_V$.

Let $P_0$ be the spot price of the securities at the starting moment $t = t_0$. The forward price $P$ of the securities at the future moment $t = t_1$ is a random variable with the mean value $\langle P \rangle > P_0$ and the standard deviation $\sigma_S$. The intrinsic yield $r_S$ determined by the forward price $P = P_0(1 + r_S)$ is a random variable with the mean value $\langle r_S \rangle$,

$$\langle r_S \rangle = \frac{\langle P \rangle}{P_0} - 1 > 0, \quad (2)$$

and the standard deviation $\sigma_{r_S} = \sigma_S / P_0$.

The discount (haircut)

$$H_c = P_0 - Q_0 > 0 \quad (3)$$

imposed by the lender to reduce the collateral risk[1], we interpret as the present price of the implicit call-option. The future revenue of the lender

$$M_g = \min\{Q_1, P\} \quad (4)$$

is a random variable with the mean value $\langle M_g \rangle < Q_1$ and the standard deviation $\sigma_{M_g}$. Define the interest rate $r_L$ of the lender,

$$r_L = \frac{M_g}{Q_0} - 1. \quad (5)$$

The amount $Q_0 = P_0(1 - h_c)$ of the lent money and the mean value of the future lender's cash in-flow $\langle M_g \rangle$ determine together the mean value of the lender's interest rate $\langle r_L \rangle$,

$$\langle r_L \rangle = \frac{\langle M_g \rangle}{Q_0} - 1 < \frac{Q_1}{Q_0} - 1 = r_R. \quad (6)$$

The value of the implicit call-option at the closing leg is the random variable

$$V = P - M_g \geq 0 \quad (7)$$

with the mean value

$$\langle V \rangle = \langle P \rangle - \langle M_g \rangle > \langle P \rangle - Q_1 > 0. \quad (8)$$

---

[1] A significant increase in the haircut index was observed during the crisis of 2007-2009 [4].

The implicit call-option present price $H_c$ (haircut) and the expected forward price $\langle P \rangle$ of the securities determine together the mean value of the yield $r_V$ of the call-option,

$$\langle r_V \rangle = \frac{\langle V \rangle}{H_c} - 1. \quad (9)$$

Manipulations with the above definitions allow us to obtain an identity relating the haircut rate $h_c = H_c / P_0$ and the mean values of the rates $r_L$, $r_S$ and $r_V$,

$$h_c(r_V - r_L) = r_S - r_L. \quad (10)$$

The definition of the mean value of the lender's return rate,

$$\langle M_g \rangle = Q_0(1 + \langle r_L \rangle) \quad (11)$$

can be used to connect the present price of the call-option and the lender's mean return. Namely, the left hand side of (11) is computed using the known probabilistic properties of $P$ and $M_g = \min\{Q_1, P\}$ at the closing moment. In the right hand side of (11), we have two unknowns, the present value $H_c = P_0 h_c$ of the implicit call-option presented in $Q_0 = P_0 - H_c$, and the expected lender's interest rate $\langle r_L \rangle$.

At this point, we observe the following opportunities:
1. Using one or another pricing model for the European call-option, e.g. the classical Black-Scholes formula, evaluate a particular value to $h_c$ and find a theoretical value to $\langle r_L \rangle$.
2. Using a reasonable valuation to $\langle r_L \rangle$, find the option pricing formula for $h_c$ alternative to the Black-Scholes formula.
3. Collect market data on $h_c$ and $r_L$ and try to construct an arbitrage strategy involving repo agreements and the traded call-options.

The implementation of the first opportunity is straightforward,

$$\langle r_L \rangle = \frac{\langle M_g \rangle}{P_0 - (H_c)_{BS}} - 1, \quad (12)$$

where $(H_c)_{BS}$ is a theoretical present price of the call-option, e.g. the Black-Scholes price.

To implement the second opportunity, it is possible to use for example a simple interest rate model based on the ergodicity property of the correlated assets,

$$\langle r \rangle - r_{rf} = C\sigma^2, \quad C = const. \quad (13)$$

Here $r_{rf}$ is the risk-free interest rate and $\sigma$ is the asset price volatility. In particular, if the asset price is strongly correlated with the security price with the return $r_S$ and the volatility $\sigma_S$, then

$$\langle r \rangle = r_{rf} + (\langle r_S \rangle - r_{rf})\frac{\sigma^2}{\sigma_S^2}. \quad (14)$$

Finally, $Q_0$ (and therefore the discount $H_c = P_0 - Q_0$) is determined by the equation

$$Q_0 = \frac{\langle M_g \rangle}{1 + r_{rf} + (\langle r_S \rangle - r_{rf})\sigma_S^{-2}\sigma_{M_g}^2}, \quad (15)$$

where $\sigma_{M_g}^2 = \langle (M_g - \langle M_g \rangle)^2 \rangle$ is the dispersion of $M_g$. Thus if one assumes that the forward price $P$ is a normally distributed random variable, then the present price $H_c$ of the call-option is a function of $Q_1$, $r_{rf}$, $\langle r_S \rangle$ and $\sigma_{M_g}$. In the example below, we verify that $H_c$ computed in this way follows numerically close to the price predicted by the Black-Scholes formula.

*Example 1.* Consider the overnight repo with a collateral portfolio of securities characterized by the annual yield $\langle r_S \rangle = 3\%$, annualized volatility $\sigma_S = 19\%$ and the spot price $P_0 = 100000$ USD. Assume that the forward price $P$ is normally distributed, has the mean value $\langle P \rangle = P_0(1 + \langle r_S \rangle / 360) = 100008,33$ USD and the standard day-to-day deviation $\sigma_S / \sqrt{360} = 1.0\%$. Then it is a straightforward computation that the repurchase price $Q_1 = (1 - 3\sigma_S / \sqrt{360})\langle P \rangle = 97003.92$ USD corresponds to almost the same mean value $\langle M_g \rangle = 97003.53$ USD and the day-to-day volatility $\sigma_{M_g} = 0.015\%$. Let the risk-free interest rate be $r_{rf} = 0\%$. Assuming that the lender accepts the above explained ergodic model of the interest rates, we find that the mean value of the lender's interest rate, $\langle r_L \rangle = r_{rf} + (\langle r_S \rangle - r_{rf})\sigma_S^{-2}\sigma_{M_g}^2 \simeq 0\%$, is almost zero. Thus the sum lent at the starting moment is $Q_0 = \langle M_g \rangle(1 + \langle r_L \rangle)^{-1} \simeq \langle M_g \rangle = 97003.53$ USD, and the discount (haircut) $H_c = P_0 - Q_0 = 2996.47$ USD (the Black-Scholes formula gives $(H_c)_{BS} = 2996.41$ USD). Finally, the corresponding repo rate is $r_R = 0.14\%$.

In the similar overnight repo with a reduced discount, letting $Q_1 = (1 - 2\sigma_S)\langle P \rangle = 98005.39$ USD, we compute

$\langle M_g \rangle = 97996.89$ USD and the day-to-day volatility $\sigma_{M_g} = 0.077\%$. Thus the mean value of the lender's interest rate is $\langle r_L \rangle = 0.018\%$ while the haircut equals $H_c = 2003.16$ USD (the Black-Scholes formula yields a little bit lower value, $(H_c)_{BS} = 2002.76$ USD). The corresponding repo rate equals $r_R = 3.1\%$.

The call-option prices $H_c$ evaluated using (14) and (15) closely follow the Black-Scholes prices in a wide range of the volatility $\sigma_S$ of the underlying securities and tend to slightly deviate from those as the strike price $Q_1$ approaches the expectation value of the forward price $\langle P \rangle$ of the securities.

### 3.2. Special Repo

The specific business models determining the behavior of the participants of the *special* repo agreement [5, 6, 8] are extremely different from those in the case of the *general* repo: the lender of funds is typically a short-term trader who lends money to borrow *specific* securities needed to maintain or expand its short position against these securities. The borrower of funds is typically a dealer or a long-term investor who owns the required securities or is able to borrow them at a third party or at the Fed's auction.

The special repurchase agreement is initiated by the lender while the borrower might not be really interesting in cash but either observes an opportunity to get the liquidity at a reduced rate or has a kind of obligation to ease the stress in the market [6].

Theoretic determination of the special repo rate is far from straightforward because the principal business idea of the trader is to make a speculative profit from the short sale at a non-equilibrium market condition.

Consideration simplifies when the lender of the specific Treasuries is the Federal Reserve while the borrower of the securities is a primary dealer.

#### 3.2.1. The Fed's Auction and the Negative Repo Rate

Interacting with the private corporations, the Fed has an upper hand and imposes particular conditions in the repo agreement to *receive* an interest. The overall repo agreement is composed as the "bonds-versus-bonds" loan with a fee determined in an auction. The fee expressed as percent per annum tends to the general repo rate, but can significantly deviate from this level [6] if the Treasuries supply is not sufficient.

The latter empirical observation can be explained revealing the standard business process of the primary dealer. The dealer lends money at the general repo rate in the general repo market and pledges the received Treasuries as the collateral for the "bonds-versus-bonds" loan. The interest gained in the general repo market is used to cover the Fed's auction fee. Thus the primary dealer maintains a neutral position on both the general repo and the Fed's auction[2]. If the demand for notes is extremely high, the dealers can become less conservative pushing the fee above the normal level.

The above plausible explanation however is not totally correct since the Fed's fee structure is more involved, see (25) and (26) below, and can not be understood without taking into account an influence of the fourth party, the dealer's client.

It is shown below (see also [6]), that the negative rate in the "bonds-versus-bonds" repo auction is closely related to the fact that the Fed never fails to deliver the agreed specific securities. The risk of not receiving the required specific securities immediately returns the special repo rate to the positive values. The mentioned above tendency of the auction fee to approach the general repo rate can be observed if the clients of the primary dealers demonstrate a high enough demand for the specific securities and by this reason agree with the almost zero special repo rates. If the market looses a particular interest to the specific securities, the Fed's auction fee drops to zero.

#### 3.2.2. Special Private Repo

Consider now the case when both parties of the agreement are private corporations. Now it is possible to figure out two kinds of fails influencing our constructions.

The first one implies that the borrower of funds (lender of the specific securities) never breaks the repurchase agreement while the lender (trader speculating against the specific securities) might hold on the short market position if the price does not fall sufficiently within the period of the repo (lender's fail).

The second kind of fails is observed when the borrower of funds does not deliver the specific securities to the lender at the starting leg receiving however the full payment (borrower's strategic fail [6]).

First consider the lender's fail case. The special repo agreement is a combined transaction involving the purchase of the securities together with the European put-option written by the borrower. A simplified version of this model is described as follows.

The dealer (borrower of funds):

1). at the starting leg, delivers the securities worth $P_0$ to the lender receiving $Q_0 = P_0(1 + h_p)$ in cash where the premium $H_p = P_0 h_p$ is the fee for "specialness";

2). at the closing leg, the dealer repurchase the same securities worth $P$ spending the agreed amount $Q_1$ in cash;

3). if the securities are not delivered by the lender at the closing day, the dealer reopens the long position at the spot

---

[2] Below, we present the formula relating the Fed's auction fee and the repo rates.

price3 $P$.

The trader (lender of money)
1). at the starting leg, lends $Q_0 = P_0(1 + h_p)$ in cash to the investor (borrower) where $H_p = P_0 h_p$ is the present price of the implicit European put-option with the strike price $Q_1$;
2). receives the specific securities worth $P_0$;
3). opens a short market position selling the securities at the spot price $P_0$;
4). at the closing leg, if the price $P$ satisfies the inequality $P < Q_1$, buys the securities at the spot price $P$ and delivers them to the borrower receiving $Q_1$ in cash (i.e. the implicit put-option is exercised);
5). if the price $P$ is too high, $P > Q_1$, then the lender extends its short market position (the implied put-option is out of money).

The special repo rate is defined as

$$r_{sR} = \frac{Q_1}{Q_0} - 1. \quad (16)$$

However computing the interest of the lender, we have to take into account its speculative profit from the short market trade, the principal reason to borrow the specific securities.

The forward price $P$ of the securities is a random variable with the mean value $\langle P \rangle$ and the volatility $\sigma_S$. Its intrinsic yield is also a random variable with the mean value $\langle r_S \rangle$ satisfying the conventional equation $\langle P \rangle = P_0(1 + \langle r_S \rangle)$.

Define the expense of the borrower at the closing leg,

$$M_s = \max\{Q_1, P\}. \quad (17)$$

The trader's net cash out-flow at the starting leg is $H_p$, while the net cash in-flow at the closing leg equals to the value $W$ of the put-option,

$$W = M_s - P = \max\{Q_1 - P, 0\}. \quad (18)$$

Thus the trader's return rate is the return rate of the put-option,

$$r_{sL} = \frac{W}{H_p} - 1. \quad (19)$$

*Example 2.* Consider the overnight special repo collateralized by the specific securities with the annual yield $\langle r_S \rangle = 3\%$, annualized volatility $\sigma_S = 19\%$ and the spot price $P_0 = 100000$ USD. Assume that the forward price $P$ is normally distributed, has the mean value $\langle P \rangle = P_0(1 + \langle r_S \rangle / 360) = 100008,33$ USD and the day-to-day volatility $\sigma_S / \sqrt{360} = 1.0\%$. Assume that the risk-free rate $r_{rf} = 0\%$. Letting the repurchase (strike) price $Q_1 = \langle P \rangle = 100008.33$ USD, the Black-Scholes formula gives the present value of the European put-option, $(H_p)_{BS} = 403.69$ USD, thus determining the lent sum $Q_0 = 100403.69$ USD. The repo rate is negative, $r_{sR} = -142\%$, in spite of the total lender's expected profit is positive, $\langle W \rangle = \langle M_s \rangle - \langle P \rangle = 399.53$ USD.

However we could not find reports on the actual market data with the values of the special repo rates comparable with the figure in Example 2. Perhaps, the dealers simply do not allow their customers (traders) to hold the short positions indefinitely long for free.

Turn now to the borrower's fail case. Similarly to the previous one, the typical borrower of funds is a dealer providing the customer (lender) with various services including the securities supply. As it is explained in [6], the market convention implies that if the borrower (lender of collateral) fails to deliver the agreed securities at the *starting* day of repo, then the agreement closes the same day but the borrower still owes the agreed interest to the lender for the full period of repo even though the lent sum is returned immediately. However if the special repo rate falls to zero, the dealer can borrow money for free while the lenders get stuck involuntary financing the dealer's short position for the period of the repo [6].

If the dealer does not own the required specific securities, then his affairs besides the special repo agreement with the trader might involve the general repo with the third parties and the special repo auction under the Fed's supervision.

For instance, the dealer starts borrowing money at the special repo rate promising to deliver the specific securities. Then it lends the borrowed cash at the general repo market receiving the general Treasuries as collateral. Both the general notes and the received interest are used, respectively, as the collateral in the "bonds-versus-bonds" loan and as a fee in the Fed's auction to borrow the specific securities. Next the dealer opens a short market position selling the specific securities and waits for a sufficient fall of the spot price. If the speculation ends successfully, the dealer closes its short position and delivers the specific securities to the client. At the closing leg, the dealer returns the borrowed money and the interest at the special rate to the trader and receives the specific securities. If the dealer's short speculation fails, nevertheless, at the closing leg, the client receives the lent money and the interest.

We interpret the dealer's fail as the execution of an American style put-option written by the lender. No doubts,

---
3 Conventionally, the delivery rescheduled to the next day at the same price [2]. However we assume that the agreement is terminated without any further consequences.

such dealer's potentially profitable but unfriendly action eventually, but not immediately, leads to the deterioration of the client base [6].

Turn now to a simplified formal description of the dealer's affairs:

1). at the starting leg, the dealer receives from a client (short-term investor or trader) $Q_0 = P_0 - H_p$ in cash; here $P_0 = p_0 N$ is the loan that has to be collateralized by $N$ specific notes at the spot price $p_0$, and $H_p = P_0 h_p$ is the haircut corresponding to the *specific* collateral;

2). lends $Q_{g0} = P_0 - H_c$ in cash and receives general securities worth $P_0$ as a collateral under a general repo agreement where $H_c = P_0 h_c$ is the haircut for the *general* collateral;

3). at the Fed's special auction, borrows $N$ specific notes of the total value $P_0 = p_0 N$ paying a fee $F_0$ in cash and leaving the general notes worth $P_0$ as collateral;

4). sells $N$ borrowed specific notes for $P_0 = p_0 N$ in the market;

5). at an intermediate leg, if the spot price $p$ of the specific notes declines sufficiently, the dealer buys $N$ specific notes in the market spending $pN = P \leq P_0 = p_0 N$ in cash;

6). delivers $N$ specific notes to the client;

7). at the closing leg, receives $N$ notes from the client and pays $Q_1 = Q_0(1 + r_{sR})$ in cash ($r_{sR}$ is the special repo rate);

8). delivers $N$ specific securities to the Fed and receives the pledged general securities;

9). receives $Q_{g1} = Q_{g0}(1 + r_R)$ in cash ($r_R$ denotes the general repo rate) and delivers the general securities to the third party.

The dealer's net cash-flow $C_B$ during the special repo period is the sum

$$C_B = C_{rf} + C_r, \qquad (20)$$

where

$$C_{rf} = Q_{g0} r_R - Q_0 r_{sR} - F_0, \qquad (21)$$

is the interest obtained from the general repo, the interest paid in the special repo and the Fed's auction fee, while

$$C_r = P_0 - P, \qquad (22)$$

is the contribution of the speculative short trade.
To avoid a financing from the other sources, the dealer has to take into account several liquidity conditions.
The first condition,

$$Q_0 - Q_{g0} - F_0 \geq 0, \qquad (23)$$

ensures the dealer's ability to finance the borrowing of $N$ specific securities at the Fed's auction using the funds provided by the client.

The second condition, $P_0 - P = (p_0 - p)N \geq 0$, allows the dealer's to purchase $N$ specific notes at the spot price $p \leq p_0$ during the special repo period.

The third condition $P_0 - P - F_0 + Q_{g0} r_R - Q_0 r_{sR} \geq 0$ is sufficient to pay the agreed sum $Q_1 = Q_0(1 + r_{sR})$ to the lender at the closing leg. However, more conservative approach requires eliminating the speculative risk from the repo obligations imposing more restrictive condition,

$$-F_0 + Q_{g0} r_R - Q_0 r_{sR} \geq 0. \qquad (24)$$

The liquidity conditions (23) and (24) together, i.e. $F_0 \leq \min\{Q_0 - Q_{g0}, Q_{g0} r_R - Q_0 r_{sR}\}$, determine the maximum fee the dealer can afford at the Fed's auction spending the client's money only. Using the relation $Q_0 = P_0(1 - h_p)$, this maximum value of the fee equals

$$\max F_0 = P_0(1 - h_p) \frac{r_R - r_{sR}}{1 + r_R}. \qquad (25)$$

The corresponding Fed's auction fee rate defined as the ratio $r_F = F_0 / P_0$ is computed as follows,

$$r_F = \frac{r_R - r_{sR}}{1 + r_{sR}}(1 - h_c) = \frac{r_R - r_{sR}}{1 + r_R}(1 - h_p) = h_c - h_p, \quad (26)$$

where $h_c = H_c / P_0$ is the haircut rate for the general collateral. This value of $F_0$ corresponds to the amount $Q_{g0}$ lent to a third party at the general repo market,

$$Q_{g0} = P_0(1 - h_p) \frac{1 + r_{sR}}{1 + r_R}, \qquad (27)$$

and since $Q_{g0} = P_0(1 - h_c)$ by definition, the above expression allows us to relate the general and special repo rates $r_R$ and $r_{sR}$ with the general and special haircut rates $h_c$ and $h_p$,

$$(1 + r_{sR})(1 - h_p) = (1 + r_R)(1 - h_c). \qquad (28)$$

As we already know, the general repo rate $r_R$ and the general repo haircut rate $h_c$ uniquely determine each other. Thus (28) can be understood as the relation between the three variables, the special repo haircut rate $h_c$ and the special and general repo rates, $r_{sR}$ and $r_R$.

Consider particular implications of the above formulas. First observe that, as soon as the Fed's auction fee is positive, $F_0 > 0$, then, according to (25) or (26), the general repo rate exceeds the special repo rate, $r_R > r_{sR}$.

Second, the relation (28) between the special and general rates implies the expression for the special haircut rate,

$$h_p = h_c \frac{1 + r_R}{1 + r_{sR}} - \frac{r_R - r_{sR}}{1 + r_{sR}}, \quad (29)$$

or, equivalently, for the special repo rate,

$$r_{sR} = \frac{r_R - h_c(1 + r_R) + h_p}{1 - h_p}. \quad (30)$$

The latter formula implies the quite strong negative value of the so-called *guaranteed-delivery* repo contracts [6]. Indeed, let $h_p = 0$ that means the absence of the put-option in the dealer's possession. Then (28) means that the Fed's fee is quite substantial (in fact, largest possible), $r_F = h_c$. At the same case, (30) yields the similar value to the special repo rate, $r_{sR} = r_R - h_c(1 + r_R)$. Indeed, the general repo rate $r_R$ is much less than the haircut rate $h_c$ for the general collateral, thus approximately, $r_{sR} \simeq -h_c$. The latter relations mean that, in the guaranteed-delivery case, the Fed's fee is covered by the dealer's client (short-term trader).

If the delivery of the specific securities is *not guaranteed* then the lender does not enter into the special repo with the negative rate, therefore $0 \le r_{sR} < r_R$, that agrees with the empirical data presented in [6]. Moreover in the situation of a stress in the market for borrowing a particular note, the special repo rate falls to zero, $r_{sR} = 0$. The corresponding specific collateral haircut rate approaches the general collateral haircut rate, $h_p = -h_c + r_R(1 - h_c) \simeq -h_c$, while the Fed's auction fee in percentage rate almost approaches the general repo rate, $r_F = r_R(1 - h_c)$, again in agreement with the empirical data [6].

In the case of absence of any particular demand to the specific notes, the Fed's auction fee is zero, $r_F = 0$, and (26) implies that the specific repo rate approaches the general one, $r_{sR} = r_R$.

## 4. Discussion

The above model of the general and special repo rates involving the call- and put-options is able to explain various phenomena observed in the repo market. This point of view also provides us with an alternative methodology of the option pricing or, conversely, of the repo rate evaluation. On the other hand, there are questions not answered yet, e.g. what is the repo rate model consistent with the Black-Scholes formula, or what are the implications of the above model on the macro-level. This moment, we are not ready to answer the first question. As to the second one, it is likely that the idea of the implicit options bought by the market participants from each other can be useful as a kind of measure of the relative degree of freedom which can be turned into a self-organizing macro-economic process.

For instance, the apparently unfair situation associated with the dealer's strategic fails, nevertheless, seems dynamically stable, cf. [6], indicating that the self-organizing potential in the system "dealer-trader" might extend beyond the conventional "service provider-customer" relationship. Indeed, one can imagine the system where the dealer, as an organizing center, identifies the goal of a speculative attack, provokes an initial stress in the market and a price movement. Looking for resources for the attack expansion and preparing the closure of his own short positions, the dealer invites the clients to join using the special repo machinery. It is known that the small traders usually wait for "trade signals" before stepping into the market, thus the *uninitiated* shocks often become unpredictable, unrecognized and even disastrous. In contrast, the trade signals reinforced using the clients' resources and the special repo machinery can benefit all the attack participants (at the expense of unaware latecomers).

## 5. Conclusions

The above presented approach to the option pricing problem involving the repo rate analysis seems fruitful and promising. Indeed, on the one hand, it sheds a new light on various faces of the repo, and, on the other hand, provides us with the new opportunities in the study of the properties of the options.

However, our simplified treatment of the business models associated with the repo agreements is in no way complete. Furthermore, the parameters of the actual repo agreements are not easy to retrieve, and our ability to proof the credibility of the above models is limited. Thus the presented remark on the repo-option relationships is rather the work in progress than the theory ready for applications.

## Acknowledgements

The author thanks Dr. O. Solomin (St. Petersburg Industrial and Construction Bank) and Dr. V. Lukianov (VTB Bank Moscow) for discussions.